\documentclass[twocolumn,,amsmath,amssymb]{revtex4}

\usepackage{graphicx}
\usepackage{dcolumn}
\usepackage{bm}
\newcommand{\etal}{\emph{et al.}}
\newcommand{\be}{\begin{equation}}
\newcommand{\ee}{\end{equation}}
\newcommand{\bfig}{\begin{figure}}
\newcommand{\efig}{\end{figure}}
\newcommand{\incl}{\includegraphics}

\begin{document}

\title{Quantum Oscillations in a Topological Insulator Bi$_2$Te$_2$Se\\
with Large Bulk Resistivity (6 $\Omega$cm)
}

\author{Jun Xiong$^1$, A. C. Petersen$^{1,2}$, Dongxia Qu$^1$, R. J. Cava$^2$ and N. P. Ong$^1$
} \affiliation{ \mbox{$^1$Department of Physics and $^2$Department of Chemistry,
Princeton University, New Jersey 08544, U.S.A.} }

\date{\today}
\begin{abstract}
We report the observation of prominent Shubnikov-de Haas oscillations in a Topological Insulator, Bi$_2$Te$_2$Se, with large bulk resistivity (6 $\Omega$cm at 4 K).  By fitting the SdH oscillations, 
we infer a large metallicity parameter $k_F\ell$ = 41, with a surface mobility ($\mu_s\sim$ 2,800 cm$^2$/Vs)
much larger than the bulk mobility ($\mu_b\sim$ 50 cm$^2$/Vs).  
The plot of the index fields $B_{\nu}$ vs. filling factor $\nu$ shows a $\frac12$-shift, consistent with massless, Dirac states.  Evidence for fractional-filling states is seen in an 11-T field.
\end{abstract}

\maketitle                   
Topological Insulators are predicted to bear current-carrying, massless, 
Dirac surface states that traverse the bulk 
energy gap~\cite{Fu07,FuKane07,Moore07,Bernevig06}.  
These unusual surface states have been observed by angle-resolved photoemission spectroscopy (ARPES)
~\cite{Hsieh08,Hsieh09,Xia09,Chen09} and scanning tunneling microscopy experiments (STM) 
~\cite{Pedram09}.  
Quantization of the Dirac states into Landau Levels has been
demonstrated in STM experiments~\cite{Hanaguri,Xue10}.  
Observation of the surface currents by transport 
has been more challenging~\cite{Check,Ando09}.  Recently, however,
encouraging progress has been achieved.  The surface SdH oscillations 
and surface mobility was measured in Bi$_2$Te$_3$ crystals~\cite{Qu}.
The existence of surface states at fractional filling in a pulsed
magnetic field were reported in (Bi,Sb)Se$_3$~\cite{Fisher}.  
Here we report the observation of prominent SdH oscillations
in crystals of Bi$_2$Te$_2$Se with very large bulk resisitivity. 
Evidence for fractional-filling states become apparent at relatively low 
magnetic fields.


\bfig[t]            
\incl[width=9cm]{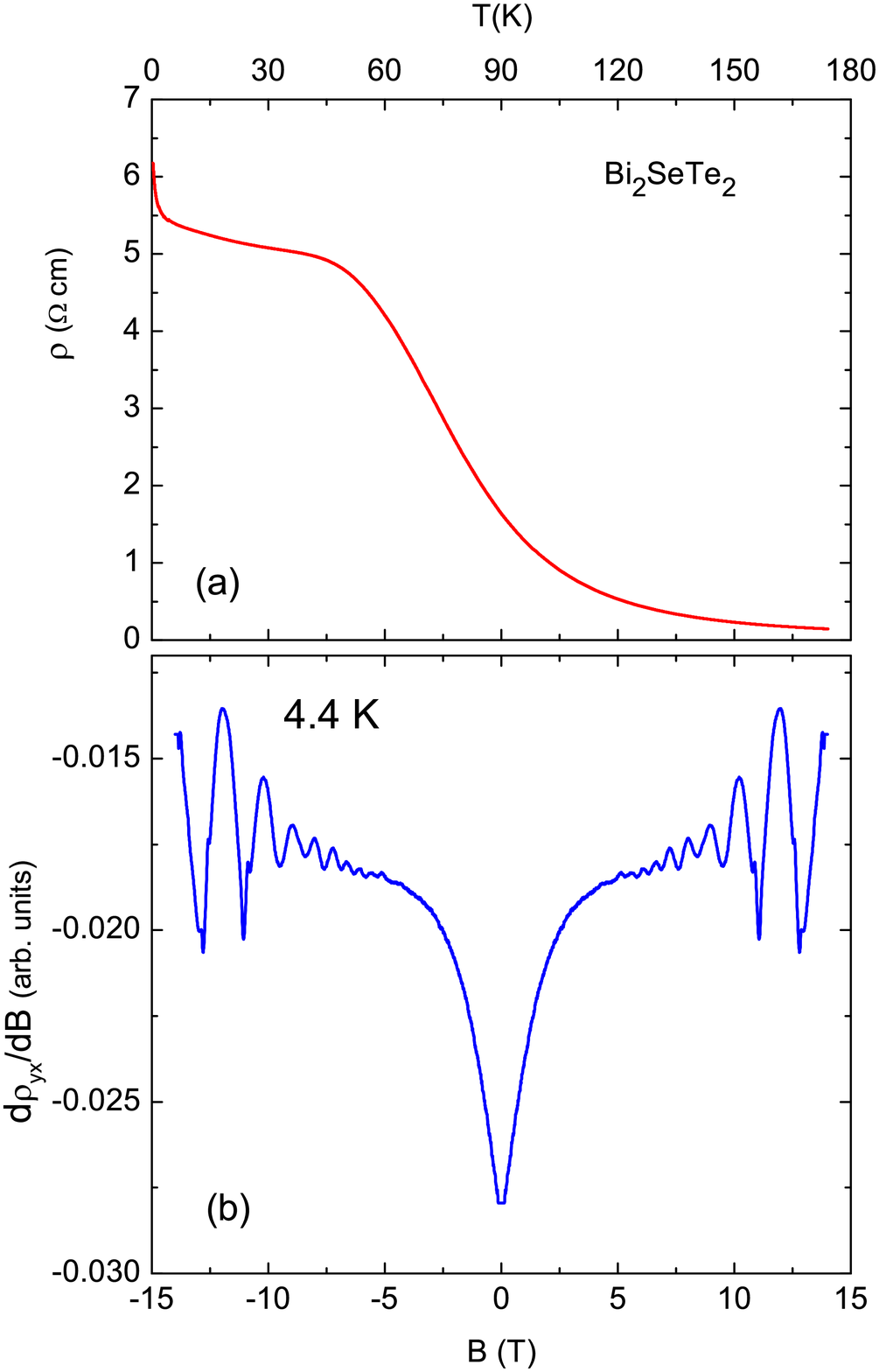} 
\caption{\label{figRvsT} (Color online) 
The resistivity of a cleaved crystal of the Topological Insulator Bi$_2$Te$_2$Se.
Panel (a) shows $\rho$ vs. $T$ measured in $B$ = 0.  Below 10 K, $\rho$ 
attains the value 5.5 $\Omega$cm, or an areal
resistance $R_{\square}$ = 400 $\Omega$.  Despite the non-metallic
value of $R_{\square}$, sizeable quantum oscillations are observed below 38 K.
Panel (b) displays the prominent SdH oscillations observed 
in the derivative $d\rho_{xy}/dB$ vs. $B$ at 4.4 K.
} 
\efig

Recently, ARPES experiments have shown that Bi$_2$Te$_2$Se displays a single 
topological surface state~\cite{Hasan}.  Independent of our experiment, 
SdH oscillations in this compound have also been reported 
by Y. Ando's group~\cite{Ando10}.  
We were motivated to grow crystals of Bi$_2$Te$_2$Se by reasoning that
Bi$_2$Se$_3$ crystals grow with a stable density of Se vacancies.
Electrons donated 
by the vacancies pin the Fermi energy $E_F$ to the conduction band,
resulting in a negative thermopower $S$ at low temperature $T$.  By contrast,
as-grown Bi$_2$Te$_3$ typically has a positive thermopower.  By growing a series of 
the hybrid semiconductor Bi$_2$Se$_{1+x}$Te$_{2-x}$, we have found that 
the low-temperature thermopower varies systematically, reflecting changes in $E_F$. 
Crystals were grown by a modified Bridgeman method from 
high purity elemental starting materials. After heating for 
one day at 850 C in a clean evacuated quartz tube, the melt 
was cooled in a temperature gradient to 500 C where it was 
left to anneal for 2 days before cooling rapidly to room temperature. 
For the composition Bi$_2$Te$_2$Se, $S$ rises to very 
large values.  With a razor blade, we cleaved thin crystals and attached
contacts using silver paint.  For the sample here, the crystal thickness $d$ = 110 $\mu$m,
while the distance between voltage leads equals 0.5 mm.  The resistivity profile
measured at $B$ = 0 is shown in Fig. \ref{figRvsT}a.
Below 40 K, the value of $\rho$ 
attains values in the range 5-6 $\Omega$cm, or $\sim$1000 times higher than in 
non-metallic Bi$_2$Te$_3$. Expressed as an areal resistance $R_{\square}= \rho/d$,
the low-$T$ resistance corresponds to $R_{\square}$ = 400 $\Omega$.  Despite the high resistance,
$\rho$ is only weakly $T$-dependent, displaying a $\log T$ increase as $T\to 0$.
The observed Hall coefficient $R_H$ below 10 K ($n$-type) implies a very small bulk
carrier concentration $n_b\sim 2.6\times 10^{16}$ cm$^{-3}$.  Combining this with
the observed $\rho$, we infer a low bulk mobility $\mu_b\sim$ 50 cm$^2$/Vs. 

Such a low $\mu_b$ should not produce SdH oscillations for $B<$14 T. Surprisingly,
however, the Hall resistivity $\rho_{yx}$ displays prominent SdH
oscillations that may be resolved up to 40 K.  To date, we have detected SdH 
oscillations in 4 crystals of Bi$_2$Te$_2$Se with 
$\rho$-$T$ profiles similar to that in Fig. \ref{figRvsT}a.
Figure \ref{figRvsT}b shows the trace of the derivative
$d\rho_{yx}/dT$ vs. $B$ at $T$ = 4.4 K.  
Independently, SdH oscillations were also observed in Bi$_2$Te$_2$Se 
by Y. Ando \etal~\cite{Ando10}. 

The surface conductance and bulk conductance act as parallel 
channels for charge transport.  The observed conductivity $\sigma_{ij}$ is
then the sum
\be
\sigma_{ij} = \sigma^b_{ij} + G^s_{ij}/d,
\label{eq:sigma}
\ee
where $\sigma^b_{ij}$ is the bulk conductivity 
and $G^s_{ij}$ the conductance matrix of the surface states.  To exploit
the additivity, we have converted the measured resistivity matrix $\rho_{ij}$ to 
the conductivity matrix $\sigma_{ij}$. 
To amplify the surface contribution to $\sigma_{xy}$, we define 
$\Delta\sigma_{xy} = \sigma_{xy} - \langle\sigma_{xy}\rangle$ 
where $\langle\sigma_{xy}\rangle$ is a smoothed background.


\bfig[t]            
\incl[width=9cm]{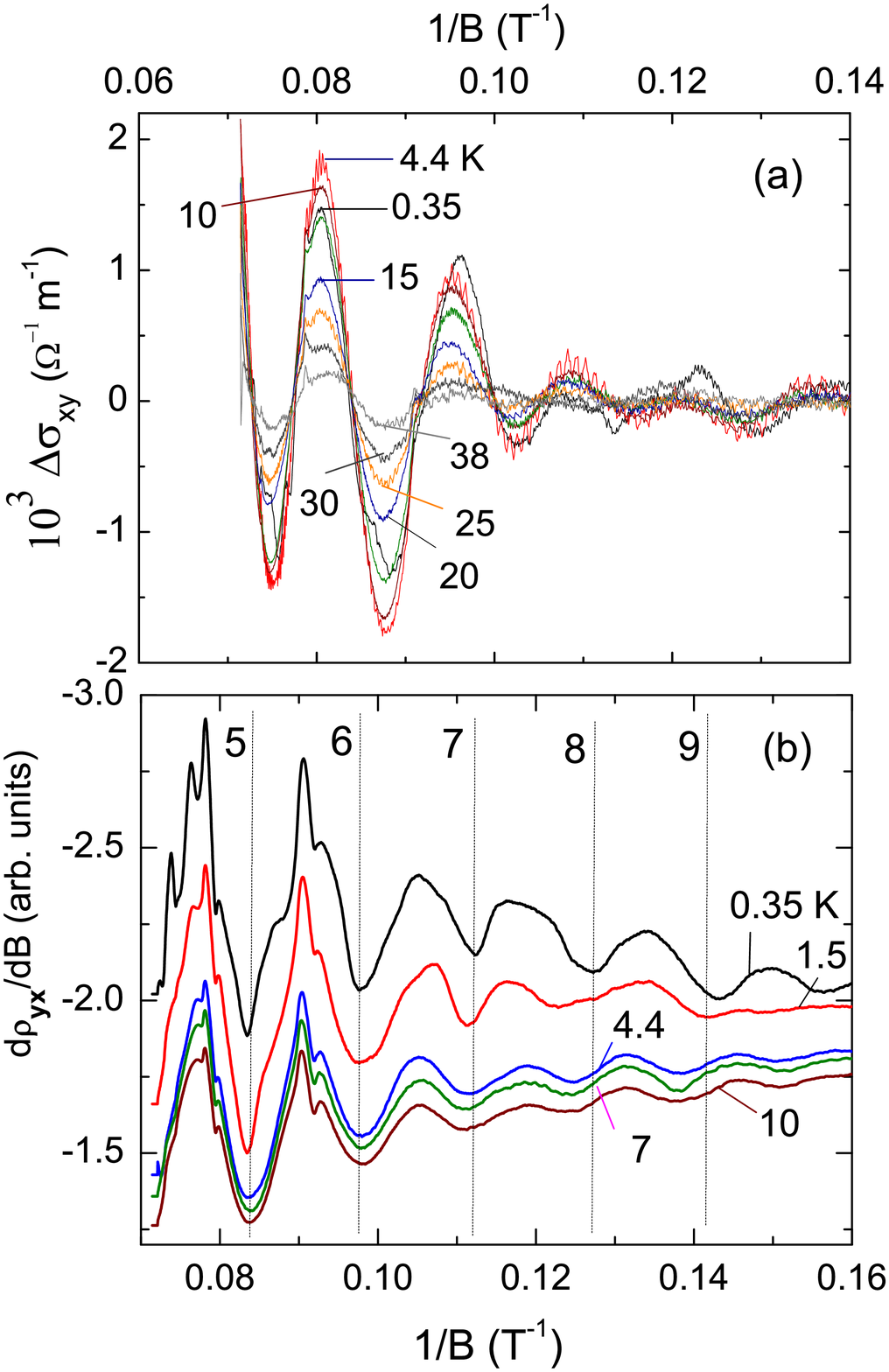}
\caption{\label{figsxy} (Color online) 
The Hall SdH oscillations versus $1/B$ at selected $T$.
In Panel (a), we have plotted the subtracted Hall conductivity
$\Delta\sigma_{xy}$ to highlight
the decrease of the amplitude over 2 decades in $T$ (0.3 to 38 K).  Panel (b)
displays traces of the Hall resistivity 
$d\rho_{yx}/dB$ at 5 temperatures.  The minima in $|d\rho_{yx}/dB|$ are used to
fix the index field $B_{\nu}$ (dashed lines with $\nu$ indicated).  
For LL with $\nu<$6, sharp structures are observed 
between integer filling.} 
\efig


\bfig[t]            
\incl[width=9cm]{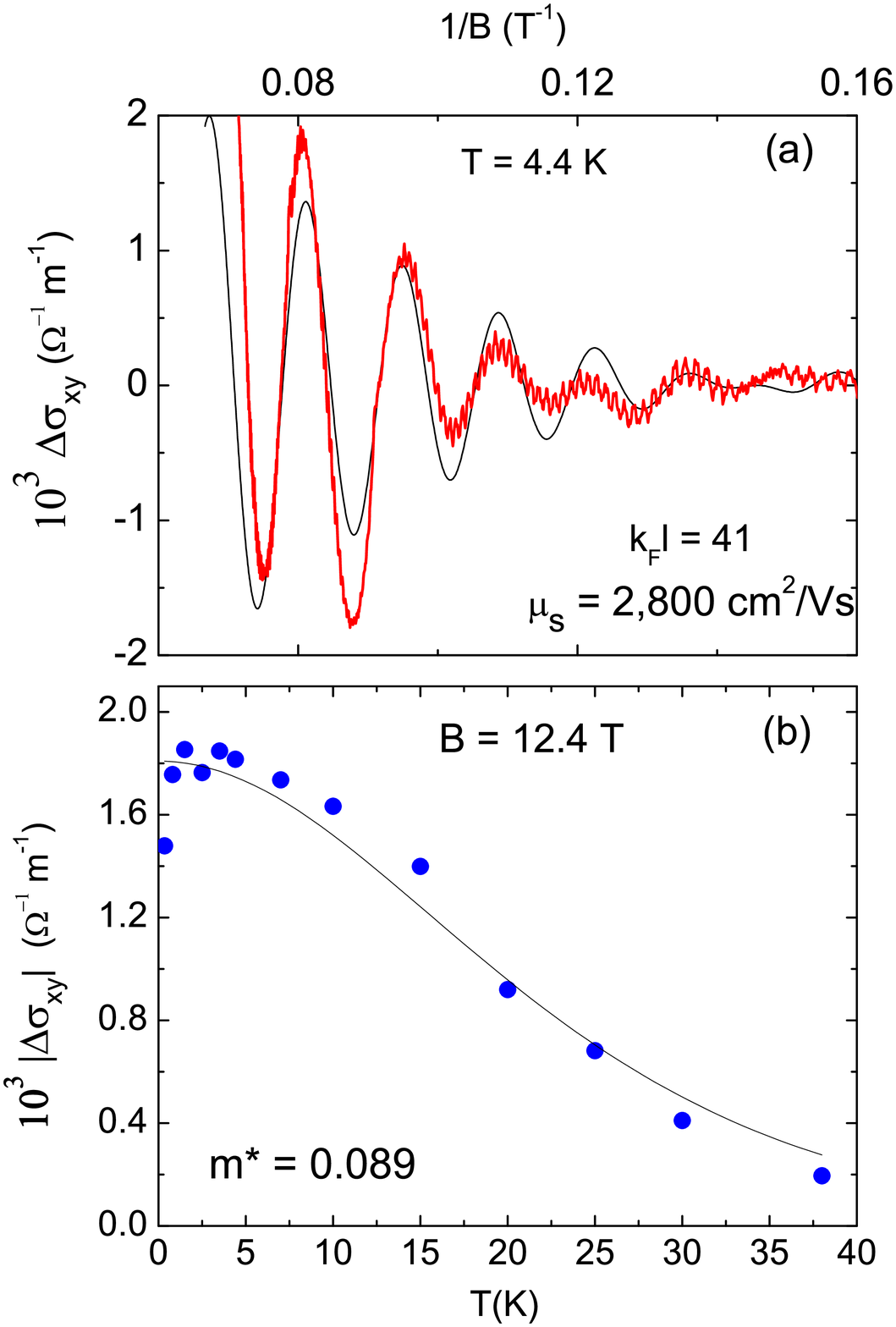} 
\caption{\label{figfit} (Color online) 
Fits of $\Delta\sigma_{xy}$ to extract mobility.  Panel (a) shows 
the fit to $\Delta\sigma_{xy}$ at $T$ = 4.4 K to Eq. \ref{eq:sdh}.
The rapid decrease of the oscillation amplitude for $1/B>$ 0.09 T$^{-1}$
reflects interference between 2 terms of equal amplitudes and 
densities ($n_{s1},\;n_{s2}$) = (1.8, 1.7)$\times 10^{12}$ cm$^{-2}$.  
The fit yields mobility $\mu$ = 2,800 cm$^2$/Vs, and $k_F\ell$ = 41.
Panel (b) shows the fit of the envelope versus $T$ with $B$ fixed at
12 T.  The fit yields $m^*$ = 0.089 $m_0$ (free mass).  With $k_F$ = 0.047 \AA$^{-1}$,
the inferred velocity $v_F$ = 6 $\times 10^5$ m/s.
} 
\efig

Figure \ref{figsxy} displays the subtracted quantity $\Delta\sigma_{xy}$
versus $1/B$ over the temperature interval 0.3 $<T<$ 38 K.
With increasing $T$, the amplitude of the SdH oscillations decreases
rapidly as $T$ increases above 10 K.

To analyze the SdH oscillations, we have fitted the curves using the standard expression~\cite{Stemmer} 
\be
\frac{\Delta\sigma_{xy}}{\sigma_{xy}} = \left(\frac{\hbar\omega_c}{2E_F}\right)^{\frac12}
\frac{\lambda}{\sinh\lambda} e^{-\lambda_D}\cos
\left[\frac{2\pi E_F}{\hbar\omega_c}+\frac{\pi}{4}\right],
\label{eq:sdh}
\ee
with $\lambda = 2\pi^2k_BT/\hbar\omega_c$ and $\lambda_D = 2\pi^2k_BT_D/\hbar\omega_c$,
where $\omega_c$ is the cyclotron frequency and 
the Dingle temperature is given by $T_D = \hbar/(2\pi k_B\tau)$,
with $\tau$ the lifetime.  Compared with
the SdH expression for the conductivity $\sigma_{xx}$, the phase $\phi$ in the 
Hall conductivity is shifted by $\pi/2$ ($-\pi/4 \to \pi/4$).
For 2D systems, we may write the SdH frequency as
$2\pi E_FB/(\hbar\omega_c)$, which simplifies to $4\pi^2\hbar n_s/e$, with the 2D
carrier density $n_s = k_F^2/4\pi$ (per spin).  
As shown in Ref.~\cite{Sharapov05}, Eq. \ref{eq:sdh} may be employed in a Dirac system
if we write the cyclotron mass as $m_c = E/v_F^2$.

We have found that the oscillations cannot be fitted using one SdH
frequency.  For $T<$ 6 K, the sharp decrease of the oscillation amplitude 
for $B^{-1}>$ 0.12 T$^{-1}$ suggests beating between 2 terms of nearly equal frequencies.
Indeed, a good fit is obtained if we add a second term identical 
to the one in Eq. \ref{eq:sdh} except for a slight difference in $n_s$.  
The measured curve of $\Delta\sigma_{xy}$ was fitted with the 2 terms
(the absolute value of the surface Hall conductance $G_{xy}$ is not known).  
There are altogether
5 adjustable parameters ($n_{si}$ and amplitude $A_i$,
with $i$ = 1,2) and $T_D$ (assumed same for both).
The best fit (Fig. \ref{figfit}a) is obtained with $A_1 = A_2$ and
and densities differing by only 5$\%$ [$(n_{s1}, n_{s2})$ = (1.79, 1.71)$\times 10^{12}$ cm$^{-2}$], corresponding to an average Fermi wavevector
$k_F$ = 0.047 \AA$^{-1}$. The fit yields $T_D$ = 8.5$\pm$1.5 K, which corresponds to a 
mean-free-path $\ell$ = 70-100 nm and a surface 
mobility $\mu_s = e\ell/\hbar k_F$ = 2,800$\pm$250 cm$^2$/Vs.
Interestingly, $\mu_s$ is strongly enhanced over $\mu_b$ (by $\sim$60).

By fitting to the decrease in amplitude with $T$ at fixed $B$ (12.4 T), we obtain
an effective mass $m^*$ = 0.089 $m_e$.  Together with $k_F$, we obtain a Fermi
velocity $v_F\sim$ 6 $\times 10^5$ m/s, higher than that in Bi$_2$Te$_3$.

The high mobility provides strong evidence that the SdH oscillations
arise from surface states.  Suppose for the sake of argument that the oscillations come from bulk states.
The SdH period must then be identified with a 3D Fermi sphere 
of radius $k_F$ = 0.047 \AA$^{-1}$, or a 3D density of 3.3$\times 10^{18}$ cm$^{-3}$.  The
inferred $\mu$ then implies a 3D resistivity $\rho_b \sim$ 0.7 m$\Omega$cm at 4.4 K.
The large discrepancy (factor of 9,000) from the observed value argues firmly against a 
bulk origin.

Hence we conclude that the SdH oscillations come from high mobility surface carriers.
The two periods likely arise from the large surfaces of the cleaved crystal that
are normal to $\bf B$.  From the inferred $n_{s1}$ and $\mu$, we find
that the conductance of each surface is $G_s = \frac12(e^2/h)k_F\ell \simeq$ 0.72 mS 
(or $R_{\square}\sim$ 1.39 k$\Omega$).


\bfig[t]            
\incl[width=9cm]{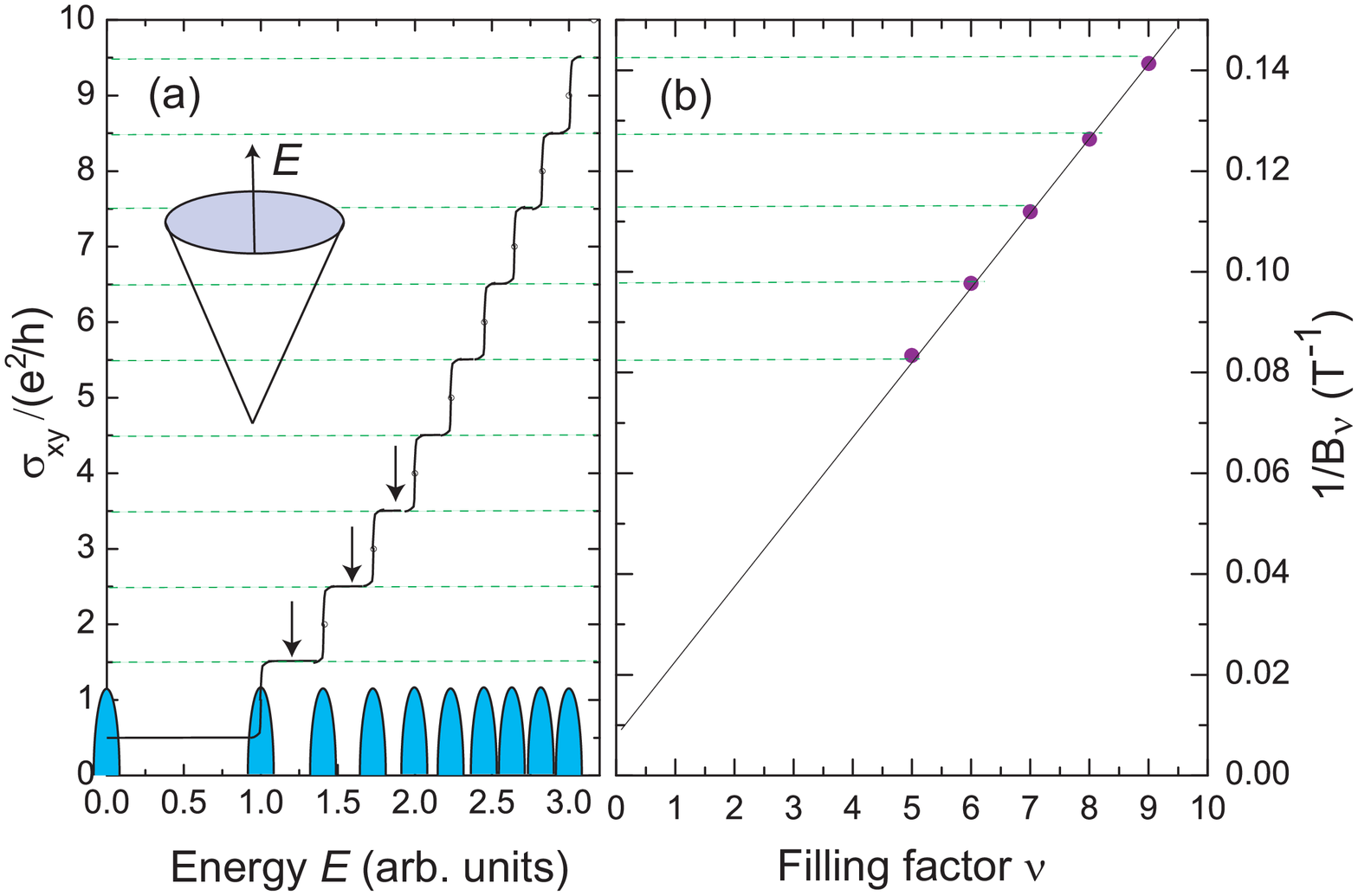} 
\caption{\label{figindex} (Color online) 
Construction used to fix the filling factor for a Dirac spectrum.  Panel (a) sketches
schematically the step-like increase of $\sigma_{xy} = (e^2/h)[\nu+\frac12]$ versus energy $E$,
where $\frac12$ arises from the $\nu$ = 0 LL at the Dirac point.
The half-ovals are broadened Landau Levels centered at $E_{\nu}=\hbar\ell_B^{-1} v_F\sqrt{2\nu}$.
$B_{\nu}$ are the fields at which $\zeta$ falls between LLs (arrows).  The inset
shows the 2D Dirac energy surface in zero $B$.  In Panel (b),
the measured values of $1/B_{\nu}$ are plotted versus filling factor $\nu$.
The 5 values of $1/B_{\nu}$ fall on a straight line 
that intercepts the $\nu$ axis at -0.55, consistent with a Dirac spectrum.  
} 
\efig

The prominence of the oscillations allows us to address the question whether
the surface states have a Dirac dispersion.  In principle, one may plot the
``index'' field $1/B_{\nu}$ versus the integers $\nu = 1,2 \cdots$,
and track the intercept in the limit $B\to\infty$.  
However, away from the quantum Hall effect (QHE) regime at low $B$, it is sometimes 
uncertain whether one should take the maxima or minima 
of $\rho_{xx}$ or $\rho_{yx}$ for the index field.
To us, the most natural choice is the field defined by the
filling factor $\nu \equiv N_e/N_{\phi}$, viz.
\be
B_{\nu} = n_s\phi_0/\nu,
\label{filling}
\ee
where $N_e$ and $N_{\phi}$ are the total number of electrons on a surface and
the number of flux quanta $\phi_0 = h/e$ piercing the surface (hereafter, we focus
on one surface, i.e. one spin degree of freedom).
Equation \ref{filling} is equivalent to having exactly $\nu$ flux quanta enclosed
in the cyclotron orbit for an electron in the LL $\nu$, viz. $B\pi k_F^2\ell_B^4 = \nu\phi_0$,
where $\ell_B = \sqrt{\hbar/eB}$ is the magnetic length.  
Hence, the chemical potential $\zeta$ lies between
2 LLs when $B = B_{\nu}$.  


\bfig[t]            
\incl[width=9cm]{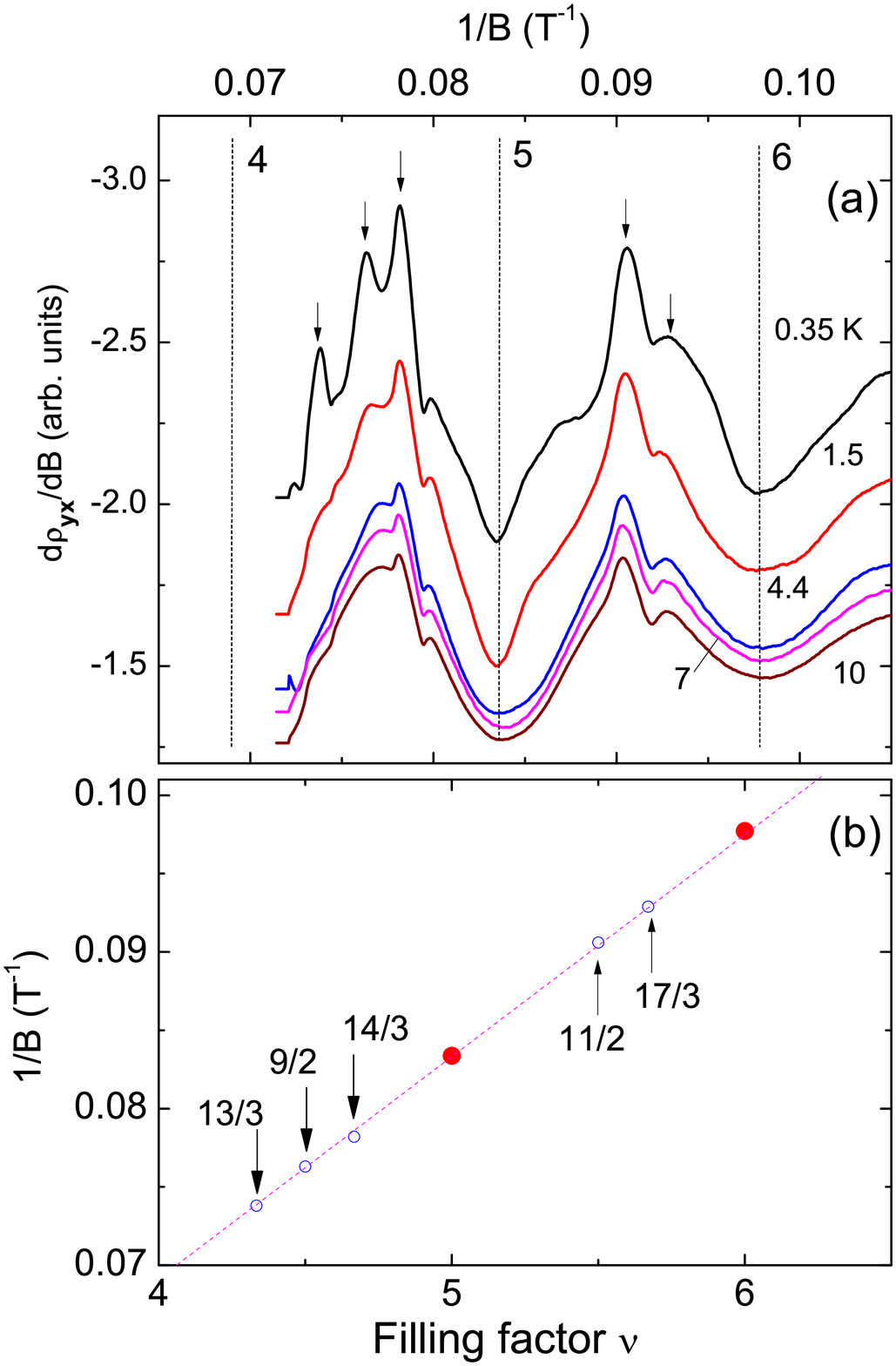} 
\caption{\label{figexpand} (Color online) 
Panel (a): Expanded view of $d\rho_{yx}/dB$ for 4$<\nu<$6 showing sharp
maxima at non-integer values of $\nu$, at $T$ from 0.35--10 K. The arrows
locate the more prominent peaks.  In Panel (b), the peak positions plotted
against $\nu$ align well with fractional values of $\nu$ (arrows mark the values
$\nu$ = 13/3, 9/2, $\cdots$, 17/3).
}
\efig

To show how the line of $1/B_{\nu}$ vs. $\nu$ relates to the Dirac dispersion, we
employ the construction in Fig. \ref{figindex}.  In Panel (a), 
we have sketched the steps of the quantized
Hall conductivity $\sigma_{xy} = (e^2/h)(\nu+\frac12)$ versus energy $E$,  
where the important shift of $\frac12$ arises from the $\nu = 0$ LL 
at the Dirac point, as in graphene.  
The half-ovals drawn on the $E$-axis represent the broadened LLs.  As the 
peaks occur at the step-edges of $\sigma_{xy}$, they also locate the maxima 
in $d\rho_{yx}/dB$.  Consequently, the minima in $|d\rho_{yx}/dB|$ locate
$B_{\nu}$, the field at which $\zeta$ lies between LLs (arrows). The $\frac12$-shift means that 
the line of $1/B_{\nu}$ vs. $\nu$ in Panel (b) does not pass through the origin (unlike
the case of a quadratic dispersion).

In Fig. \ref{figindex}b, we have plotted the measured $1/B_{\nu}$ versus $\nu$.
With just one vertical scale adjustment, we can align the 5 data points to the steps
in Panel (a).  At our highest $B$, the filling factor is $\nu$ = 5.
More significantly, the line through the data in (b) intercepts
the $\nu$ axis at $\nu$ = -0.55 instead of 0.  Hence our results are consistent
with the $\frac12$ shift expected from a Dirac spectrum.  

Recently, Fisher \etal~observed fractional-filling states in (Bi,Sb)Se3
in very intense fields ($>$50 T).  In our experiment, evidence for fractional-filling 
begin to emerge at much lower fields (11 T).  As shown in Fig. \ref{figexpand}a,
an intriguing array of sharp \emph{maxima} in $d\rho_{yx}/dB$ (arrows) 
is apparent for 4$<\nu <$6. The peaks become weaker as $T$ increases from 0.35 K, but some 
are still resolved at 10 K.  Interestingly, the peak positions 
align well with fractional values of $\nu$, as shown in Fig. \ref{figexpand}b.

As in the fractional QHE regime in GaAs and graphene, we interpret the peaks 
as evidence for many-body states that are stabilized at fractional $\nu$.  
However, there are several puzzling features (when compared with the 
standard FQHE phenomenology).  First, in Fig. \ref{figexpand}a, we find that the maxima
of $|d\rho_{yx}/dB|$ locate the fractional values of $\nu$ whereas $B_{\nu}$ for integer $\nu$ is
fixed by the minima.  We do not have an explanation for this inversion.
Second, the fractional-filling peaks are observed at fairly large $\nu$, whereas in GaAs,
they are difficult to resolve for $\nu>$2, despite the much higher $\mu_s$ in GaAs.
Finally, unlike in GaAs, the fractions corresponding to $\frac12$ 
are more prominent here (particularly 11/2)
than those corresponding to $\frac13$ or $\frac23$.  
In this system, we may incorporate featurs such as the linear Dirac dispersion, 
strong spin-orbit interaction, and strong
suppression of 2$k_F$ scattering to understand the states at fractional filling.

We acknowledge support from the National Science Foundation under grant
DMR 0819860 and from the Nano Electronics Research Corporation (Award 2010-NE-2010G). 
We thank Liang Fu, Duncan Haldane and Andrei Bernevig for explaining the filling of LLs in the 2D Dirac system.


%

%
\end{document}